\documentclass[preprint,preprintnumbers,amsmath,amssymb]{revtex4}
\usepackage{graphicx}% Include figure files
\usepackage{dcolumn}% Align table columns on decimal point
\usepackage{bm}% bold math
\usepackage{amsmath,bm}
\usepackage{xcolor,graphicx,float}
\usepackage{subfigure}
\usepackage{amssymb}
\usepackage{natbib}

\begin{document}

\title{Decoherence of Dirac-particle quantumness for fermionic fields in a dilatonic black hole}

\author{Chengjun Yao}
\author{Yating Shao}
\author{Kai Yan}
\author{Yinzhong Wu}
\affiliation{School of Physical Science and Technology, Suzhou University of Science and Technology, Suzhou,Jiangsu 215009, People's Republic of China}
\author{Xiang Hao}
\altaffiliation{Corresponding author}
\email{xhao@mail.usts.edu.cn}

\affiliation{School of Physical Science and Technology, Suzhou University of Science and Technology, Suzhou,Jiangsu 215009, People's Republic of China}

\affiliation{Pacific Institute of Theoretical Physics, Department of Physics and Astronomy,
\\University of British Columbia, 6224 Agriculture Rd., Vancouver B.C., Canada V6T 1Z1.}

\begin{abstract}

The quantumness of Dirac paticles for quantized fields in a dilatonic black hole is estimated by means of quantum channel. We develop a general Bloch vector representation of quantum channel in black hole spacetimes beyond single mode approximation. The nonclassicality of Dirac particles can be measured by the minimization of quantum coherence over all orthonormal basis sets. The quantumness of the channel decreases as the dilaton parameter increases. The interplay between the external reservoir noise and dilaton black hole on the dynamical behavior of quantum coherence and steerability is investigated in the Pauli basis. The external environment is modelled by a random telegraph noise channel. The monotonous decay of quantum nonlocality occurs in the weak coupling case. The degradation and revival of quantum nonlocality are observed in the strong coupling condition. It is found that quantum fluctuation effects of the external reservoir can protect quantum coherence and steerability from the information loss of the black hole.

\vspace{1.6cm} Keywords: dilatonic black hole, quantum channel, quantum coherence, quantum steering

\end{abstract}

\maketitle
\section{Introduction}

The observation of gravitational waves and confirmation of black holes \cite{Klaer2017} have received much attention. The quantum property of field modes near the event horizon plays a prominent role in the study of black hole thermodynamics \cite{Huang2021}. The relativistic effect on quantum dynamics in accelation frames \cite{Giacomini2019} and vacuum fluctuation in black hole spacetimes \cite{Wang2016} have been studied in the field of relativistic quantum information \cite{Peres2004}. As is well known, Unruh acceleration effects \cite{Unruh1976,Banerjee2017} and gravitational fields \cite{Hawking1975,He2016} lead to information loss of particles or antiparticles in a relativistic framework. Recent studies have demonstrated that the geometric inflation and shadows have impacts on the nonclassicality of quasibound states of black holes \cite{Edelstein2020}. From the viewpoint of quantum nonlocality \cite{Nielsen2011}, quantumness of scalar and Dirac fields in the black holes can be quantified by quantum entanglement \cite{Martin2012} and quantum correlation\cite{Wen2020}. The relativistic metrology based on quantum Fisher information was also applied to the analysis of vacuum structures in curved spacetimes \cite{Huang2019,Jin2020}. The class of dilatonic black holes has become an interesting focus in astrophysical black holes \cite{Anabalon2013,Amarilla2013,Badia2020,Kumar2020}. The superradiant instability and scattering properties of dilatonic black holes are studied in the references of \cite{Bernard2017,Huang2020}. In the low energy limit of string theory, the static spherical charged dilatonic black hole solution was provided by Gibbons and Maeda \cite{Gibbons1988}, independently by Garfinkle, Horowitz, and Strominger \cite{Garfinkle1991}($\mathrm{GMGHS}$). The study of dilatonic black holes contributes to a deeper understanding of quantum gravity \cite{Howl2021}. It is of a great interest to explore quantumness of dilatonic black holes in the combination of quantum information and general relativity.

On the other hand, quantum fluctuations in some curved spacetimes can be investigated in the context of open quantum system approach \cite{Yu2011}.  As one feasible method, quantum channel can be used to characterize dynamics of quantum systems interacting with their reservoirs \cite{Scully1997,Barnett1997}. The random telegraph noise channel \cite{Daffer2004,Benedetti2013}, for instance, can describe quantum dissipation from a structured reservoir with a fluctuation rate. Different types of environments bring out diverse decoherence. In the background of black holes, we expect to find a unified quantum channel which discovers the connection between gravitation and decoherence. The quantum channel can be represented as the evolution of Bloch states for quantized fields in relativistic spacetimes. The quantum resource theory \cite{Baumgratz2014} provides an efficient way to measure quantumness of channels \cite{Shahbeigi2018}. In fact, no feasible localized detector models can be applied to the measurement of quantum effects of fields inside the event horizon of a black hole. Therefore, it is advantageous to probe quantum nonlocality near the outside region of the event horizon of black holes. We assume that the two observers can detect local field modes. After both of them share a quantum correlated initial state, one stays stationary at an asymptotically flat region and experiences quantum dissipation from an external reservoir, while the other moves with uniform acceleration and hovers near the event horizon of the dilatonic black hole. In the frame of quantum channel, quantum coherence and quantum steering \cite{Jevtic2014,Costa2016} are used to estimate the dynamical behavior of open Dirac particles.

The paper is organized as follows. In the next section, the quantized fields of Dirac particles and antiparticles are discussed beyond single mode approximation. We suggest the generalized representation of quantum channel in the form of Bloch vectors. The available estimation of quantumness of the channel is obtained by means of quantum coherence. The relativistic effects from the dilatonic parameter and Dirac mode approximation are considered. In Sec. III, the interplay between the external environment and dilatonic black hole spacetime is analyzed in terms of quantum coherence and quantum steering. We put forward a unified method of quantum decoherence in the Pauli basis. We will discuss our conclusion in the last section.

\section{Quantum channel in the dilatonic black hole}

In accordance with string theory, the dilatonic black holes are formed by gravitational systems coupled to Maxwell and dilaton fields. The action of the four-dimensional dilaton theory is written as \cite{Gibbons1988,Garfinkle1991}
\begin{equation}
\label{eq:(1)}
\mathcal{S}=\int \mathrm{d}^4x \sqrt{-g}[R-2(\nabla \phi)^2-e^{-2a\phi}F_{\mu \nu}F^{\mu \nu}],
\end{equation}
where $\phi$ represents the dilaton and $F_{\mu \nu}=\partial_{\mu}A_{\nu}-\partial_{\nu}A_{\mu}$ is the $\mathrm{Maxwell}$ tensor. The coupling parameter between the dilaton and $\mathrm{Maxwell}$ field is given by $a=1$ which corresponds to the $\mathrm{GMGHS}$ black hole. The line element of this spherical dilatonic black hole is given by
\begin{equation}
\label{eq:(2)}
\mathrm{d}s^2=-\big(\frac{r-2M}{r-2D}\big)\mathrm{d}t^2+\big(\frac{r-2M}{r-2D}\big)^{-1}\mathrm{d}r^2+r(r-2D)(\mathrm{d}\theta^2+\sin^2\theta \; \mathrm{d}\phi^2).
\end{equation}
The natural units $\hbar=G=c=\kappa_B=1$ are used. The parameters $M$ and $D$ are the mass of the black hole and dilaton respectively. For a charged dilatonic black hole, $D=\frac {Q^2}{2M}$ where $Q$ denotes the charge. The term $e^{-2\phi}=e^{-2\phi_0}(1-\frac {2D}{r})$. When the dilaton value at spacelike infinity $\phi_0=0$, the case implies an asymptotic flat manifold. The event horizon of $\mathrm{GMGHS}$ black hole is located at $r_{+}=2M$. The area of the sphere goes to zero when $r=r_{-}=2D$ and the surface is singular. The quantized Dirac fields in the dilatonic black hole satisfy the Dirac equation which is given by $\gamma^{\alpha} e_{a}^{\mu}(\partial_\mu+\Gamma_\mu)\Psi=0$ \cite{Wang2010}. $\gamma^{\alpha}$ is the Dirac matrix, $e_a^\mu$ corresponds to the inverse of the tetrad $e^{\mu}_{a}$ and $\Gamma_{\mu}$ denotes the spin connection coefficient. By solving the Dirac equation, we can obtain the positive frequency outgoing solutions outside region and inside region of the event horizon, $\Psi_{\iota,\boldsymbol{k}}^{+}=\Re e^{\mp i\omega \mathcal{U}}$ where $\iota=(\mathrm{out},\mathrm{in})$ represents the two regions, $\boldsymbol{k}$ denotes a variable that labels the field mode, $\Re$ is a $4$-component Dirac spinor and $\omega$ is a monochromatic frequency of the Dirac field. The parameter $\mathcal{U}=t-2(M-D)\ln[(r-2M)/(2M-2D)]$ is the retarded time.

In terms of the complete orthogonal basis $\psi_{\iota,\boldsymbol{k}}^{\pm}$, the Dirac field can be expanded as
\begin{equation}
\label{eq:(3)}
\Psi=\sum_{\iota=(\mathrm{out},\mathrm{in})}\int \mathrm{d} \boldsymbol{k} (\hat{a}_{\boldsymbol{k}}^{\iota} \Psi_{\iota,\boldsymbol{k}}^{+}+\hat{b}_{\boldsymbol{k}}^{\iota \dag}\Psi_{\iota,\boldsymbol{k}}^{-}),
\end{equation}
where $\hat{a}_{\bm k}^{\iota}$ and $\hat{b}_{\bm k}^{\iota \dag}$ correspond to the fermion annihilation and antifermion creation operators acting on the state of the $\iota$ region. Making the analytic continuations for $\Psi_{\iota,\boldsymbol{k}}^{+}$, we can quantize Dirac fields as the combination of Kruskal modes according to the suggestion of Domour-Ruffini \cite{Damoar1976}. The quantization of Dirac fields beyond single mode approximation \cite{JWang2010} can be taken into account. Using the Bogoliubov transformations between the creation and annihilation operators in two kinds of coordinates, the operators for mode $\bm k$ are expressed as \cite{Bruschi2010,Friis2011}
\begin{align}
\label{eq:(4)}
\hat{d}_{R,\boldsymbol{k}} &\;=\; \cos r \hat{a}_{\bm k}^{\mathrm{out}}-\sin r \hat{b}_{-\bm k}^{\mathrm{in}\dag}   \nonumber \\
\hat{d}_{L,\boldsymbol{k}} &\;=\; \cos r \hat{a}_{\bm k}^{\mathrm{in}}-\sin r \hat{b}_{-\bm k}^{\mathrm{out}\dag},
\end{align}
where $\cos r =c= [e^{-8\pi \omega(M-D)}+1]^{-\frac 12} $ and $\sin r =s= [e^{8\pi \omega(M-D)}+1]^{-\frac 12} $. $\hat{d}_{R(L),\boldsymbol{k}}$ denotes the annihilation operator for the right ($\mathrm{R}$) or left ($\mathrm{L}$) mode. Assuming that $|0_{\bm k}\rangle_{R(L)}$ is annihilated by the operator $\hat{d}_{R(L),\boldsymbol{k}}$, the Kruskal vacuum for mode $\bm k$ is defined as $|0\rangle_{K}=|0_{\bm k}\rangle_{R}|0_{\bm k}\rangle_{L}=c^2|0000\rangle-sc|0011\rangle-sc|1100\rangle-s^2|1111\rangle$. Here, $|mn\tilde{m}\tilde{n}\rangle=|m_{\bm k}\rangle^{+}_{\mathrm{out}}|n_{-\bm k}\rangle^{-}_{\mathrm{in}}|\tilde{m}_{\bm -k}\rangle^{-}_{\mathrm{out}}|\tilde{n}_{\bm k}\rangle^{+}_{\mathrm{in}}$ and the superscripts $\{ +, -\}$ represent the particle and antiparticle. When the observer is located outside the event horizon, the Hawking radiation spectrum with a thermal Fermi-Dirac distribution can be obtained by $\langle 0 |\hat{a}_{\bm k}^{\mathrm{out}\dag}\hat{a}_{\bm k}^{\mathrm{out}}|0\rangle=\frac 1{e^{\omega/T}+1}$ where $T=\frac 1{8\pi(M-D)}$ is the Hawking temperature \cite{Hawking1975}. Because of the Pauli exclusion principle, only the first excited state for particle mode is obtained by
\begin{align}
\label{eq:(5)}
|1\rangle_{K}^{+} &=\;[q_{R}(\hat{d}_{R,\boldsymbol{k}}^{\dag}\otimes I_L)+q_L(I_R\otimes \hat{d}_{L,\boldsymbol{k}}^{\dag})]|0\rangle_{K}\nonumber \\
&=\; q_R(c|1000\rangle-s|1011\rangle)+q_L(s|1101\rangle+c|0001\rangle),
\end{align}
with $|q_R|^2+|q_L|^2=1$. The physical interpretation on the operators of Eq. (4) and states in Eq. (5) beyond single mode approximation was presented \cite{Wang2014}. The creation operator $\hat{d}_{R,\boldsymbol{k}}^{\dag}$ implies the creation of a fermion in the exterior vacuum and an antifermion in the interior vacuum of the black hole. The spontaneous creation of particles and antiparticles results in Hawking radiation with the total probability $|q_R|^2+|q_L|^2=1$. When the parameter $|q_R|=1$, all the fermions propagate toward the outside of event horizon and all the antifermions move to the interior region. Similarly, the case of $|q_L|=1$ represents that all the antifermions radiate toward the exterior region.

Since the exterior region is causally disconnected from the interior region in the dilatonic black hole, we can obtain the physical accessible outside states by tracing over the states of the interior region. The whole process of information loss can be characterized by a quantum channel. In the open quantum system approach, the dynamical map for the system of interest can be obtained by tracing over all the degrees of freedom of the other environment. With respect to the black hole spacetime, the outside states of all particles or antiparticles can be considered as the open system of interest. Thus, the quantum channel for Dirac particles in the outside region of the event horizon can be given by
\begin{align}
\label{eq:(6)}
\boldsymbol  \varepsilon^{+}_{\mathrm{out}}(|0\rangle_{K}\langle0|)&\;=\; c^2 |0\rangle^{+}_{\mathrm{out}}\langle0| +s^2 |1\rangle^{+}_{\mathrm{out}}\langle1|\nonumber \\
\boldsymbol  \varepsilon^{+}_{\mathrm{out}}(|0\rangle_{K}\langle1|)&\;=\; q_R^{\ast}\cdot c |0\rangle^{+}_{\mathrm{out}}\langle1|  \nonumber \\
\boldsymbol  \varepsilon^{+}_{\mathrm{out}}(|1\rangle_{K}\langle0|)&\;=\; q_R \cdot c  |1\rangle^{+}_{\mathrm{out}}\langle0|  \nonumber \\
\boldsymbol  \varepsilon^{+}_{\mathrm{out}}(|1\rangle_{K}\langle1|)&\;=\;  |q_L|^2c^2 |0\rangle^{+}_{\mathrm{out}}\langle0| +(|q_R|^2+|q_L|^2s^2) |1\rangle^{+}_{\mathrm{out}}\langle1|,
\end{align}
where the partial trace over all the states of the interior region and the antiparticle states of the outside is carried out. For single particle state, the density matrix of the state $\rho$ can be expressed by the Bloch vector $\bm \upsilon$, i.e., $\rho=\frac{1}{2}(\mathbb{I}+\sum_{\mu=1}^{3}\upsilon_{\mu}\sigma_{\mu})$ where $\upsilon_\mu=\mathrm{Tr}(\rho\sigma_\mu), \quad (\mu=1,2,3)$ and $\sigma_{\mu}(\mu=1,2,3)$ denote the three Pauli operators. In the pauli basis of $\{ \sigma_0=\mathbb{I},\sigma_{\mu}(\mu=1,2,3)\}$, the dynamical process from an arbitrary initial particle vector $\boldsymbol {X}=(1,\boldsymbol {\upsilon}^{(0)})$ to the evolved one  $\boldsymbol {Y}=(1,\boldsymbol {\upsilon})$ is written as
\begin{equation}
\label{eq:(7)}
\boldsymbol  Y=\boldsymbol \hat{\Phi}^{+}_{\mathrm{out}} \cdot \boldsymbol  X=\Big(\begin{matrix} 1&\bm {0}^{\mathrm{T}} \\ \boldsymbol \lambda^{+}_{\mathrm{out}} &\boldsymbol {\Gamma }^{+}_{\mathrm{out}}\end{matrix}\Big)\Big(\begin{matrix} 1\\ \boldsymbol \upsilon^{(0)}\end{matrix}\Big),
\end{equation}
where $\boldsymbol  {0}^{\mathrm{T}}=(0,0,0)$ and the superscript $\mathrm{T}$ denotes the transpose. The quantum channel $\boldsymbol  \varepsilon^{+}_{\mathrm{out}}(\cdot)$ is equivalent to the action operator $\boldsymbol \hat{\Phi}$  which is composed of the mapping vector and mapping matrix. Here, $\lambda_\mu(\mu=1,2,3)=\mathrm{Tr}[\frac{\sigma_\mu}{2}\cdot\boldsymbol {\varepsilon}^{+}_{\mathrm{out}}(\mathbb{I})]$. The elements of the matrix $\boldsymbol {\Gamma }$ are obtained by $\Gamma_{\mu,\nu}(\mu,\nu=1,2,3)=\mathrm{Tr}[\frac{\sigma_\mu}{2}\cdot\boldsymbol {\varepsilon}^{+}_{\mathrm{out}}(\sigma_\nu)]$. By using Eq. (6), we obtain the channel matrix and vector as
\begin{equation}
\label{eq:(8)}
\boldsymbol {\Gamma }^{+}_{\mathrm{out}}=\left(\begin{matrix} \mathrm{Re}(q_R)\cdot c& -\mathrm{Im}(q_R)\cdot c & 0 \\ \mathrm{Im}(q_R)\cdot c &  \mathrm{Re}(q_R)\cdot c & 0 \\ 0 & 0 & |q_R|^2 \cdot c^2 \end{matrix}\right), \quad \boldsymbol \lambda^{+}_{\mathrm{out}}= \left(\begin{matrix} 0\\0\\(1-|q_R|^2)c^2-s^2 \end{matrix}\right).
\end{equation}
We consider a real parameter, i.e., $\mathrm{Re}(q_R)=q_R\in[0,1]$ and $\mathrm{Im}(q_R)=0$ in this case. The evolved particle state is expressed as $\boldsymbol {\upsilon}=\boldsymbol {\Gamma }^{+}_{\mathrm{out}}\cdot \boldsymbol {\upsilon}^{(0)}+ \boldsymbol \lambda^{+}_{\mathrm{out}}$ in the form of Bloch vectors. It is known that a geometric picture of an arbitrary pure state is a symmetric sphere which corresponds to a Bloch vector $\boldsymbol {\upsilon}^{(0)}=(\cos \phi \sin \theta, \sin \phi \sin \theta, \cos \theta)^{\mathrm{T}}, \; \theta \in[0,\pi), \phi \in[0,2\pi)$ without information loss. If the quantum channel of Eq. (8) acts on a Dirac qubit, the sphere deformation is depicted in Figure 1(a). Under the condition of the small values of $q_R$, the Bloch sphere for Dirac particle states shrinks drastically. The geometrical contraction can verify the fact about the loss of quantum information induced by quantum fluctuations near the event horizon of the black hole.

Moreover, we try to quantify the quantumness of the channel which describes the thermal radiation of particles near the event horizon of the dilatonic black hole. The squared $l_1$ norm can be used to measure quantum coherence of a particle state. The quantity of quantum coherence is given by $C(\rho)=\sum_{i\neq j}|\varrho_{ij}|$ where $\varrho_{ij}$ is the element of the state density matrix. When a particle state is subjected to the action from quantum channel, the decay of quantum coherence depends both on the channel and incoherent basis. The nonunitary evolution of states addresses the question of how much quantumness a channel can preserve. By averaging on quantum coherence of all states and minimizing over all orthonormal basis sets \cite{Shahbeigi2018}, the measure of nonclassicality of the quantum channel $\boldsymbol  \varepsilon^{+}_{\mathrm{out}}$ can be obtained as
\begin{equation}
\label{eq:(9)}
Q(\boldsymbol  \varepsilon^{+}_{\mathrm{out}})=\min_{\bm n}[\mathrm{Tr}(\mathcal{\bm M})-\bm n^{\mathrm{T}} \mathcal{\bm M}\bm n ],
\end{equation}
where $\mathcal{\bm M}=\frac 12(\boldsymbol {\Gamma }^{+}_{\mathrm{out}}\boldsymbol {\Gamma }^{+\mathrm{T}}_{\mathrm{out}}+5\boldsymbol \lambda^{+}_{\mathrm{out}}\boldsymbol \lambda^{+\mathrm{T}}_{\mathrm{out}})$ is determined by the mapping vector and matrix. The minimization is achieved when $\bm n$ is chosen to be an eigenvector of the matrix $\mathcal{\bm M}$ with the largest eigenvalue. Therefore, the quantumness of the channel is calculated by $Q(\boldsymbol  \varepsilon^{+}_{\mathrm{out}})=\mathrm{Tr}(\mathcal{\bm M})-\max (m_1,m_2,m_3)$ where $m_1,m_2,m_3$ are the eigenvalues of the matrix. The channel quantumness from quantum fluctuations near the event hozion is related to the dilatonic parameter $D$ and the probability $q_R$ is illustrated in Figure 1(b). The increase of the dilationic parameter $D$ will diminish the quantumness. When $q_R \rightarrow 1$, the quantity of the nonclassicality in the black hole can keep a large value. If the dilatonic parameter $D>0.8$, the quantumness declines rapidly. When an observer hovers near the event horizon of the black hole, the detection of Dirac particle states from the outside region can be applied to the measure of quantumness.

\section{Relativistic decoherence with external noise}

The Dirac particles escaping from the outside region may experience external noisy environments. When one observer detects states of open Dirac particles, the effects from external noises need be considered. It is necessary to study the impacts from external noises on quantum nonlocality in the background of black hole spacetime. In a protocol, two observers can detect local field modes of Dirac particles. Both of them share an initial state of two Dirac particles. One observer stays stationary at an asymptotically flat region and the local Dirac particle $A$ suffers from an external reservoir. While the other detects the state of Dirac particle $B$ near the event horizon of the black hole. In terms of quantum channels, the two Dirac particles experience two independent channels of $\boldsymbol  \varepsilon_{A}$ and $\boldsymbol  \varepsilon_{B}$  respectively. In general, an arbitrary state for two Dirac particles $A$ and $B$ can be written as $\rho_{AB}^{(0)}=\frac 14 \sum_{\mu,\nu=0}^{3}\Theta_{\mu\nu}^{(0)}(\sigma_{\mu,A}\otimes \sigma_{\nu,B})$. Here, the matrix element $\Theta_{\mu\nu}^{(0)}=\mathrm{Tr}(\rho_{AB}^{(0)}\sigma_{\mu,A}\otimes \sigma_{\nu,B})$ is obtained. In the pauli basis, we can obtain the equivalent form of the initial state by using the matrix
\begin{align}
\label{eq:(10)}
\boldsymbol{\Theta}^{(0)}=\left( \begin{matrix} 1 & \boldsymbol \upsilon_{B}^{(0)\mathrm{T}} \\ \boldsymbol \upsilon_{A}^{(0)} & \boldsymbol T^{(0)} \end{matrix}\right),
\end{align}
where $\boldsymbol \upsilon_{A(B)}^{(0)}$ denotes the initial Bloch vector of the particle $A$ or $B$. The matrix $\bm T^{(0)}$ implies the correlation information \cite{Horodecki1996,Luo2003}. Each quantum channel is determined by the mapping vector $\boldsymbol \lambda$ and matrix $\boldsymbol  \Gamma$. In the Bloch vector representation, the transformation of quantum channel is given by $\boldsymbol  \varepsilon: \boldsymbol {\upsilon}=\boldsymbol {\Gamma }\cdot \boldsymbol {\upsilon}^{(0)}+ \boldsymbol \lambda $. In the context of open quantum systems, the evolved state is obtained by $\rho_{AB}=\boldsymbol  \varepsilon_{A} \otimes \boldsymbol  \varepsilon_{B}(\rho_{AB}^{(0)})$. We can expand $\rho_{AB}$ as
\begin{equation}
\label{eq:(11)}
\rho_{AB}=\frac 14 \sum_{\mu,\nu=0}^{3}\Theta_{\mu\nu}^{(0)}[\boldsymbol {\varepsilon}_{A}(\sigma_{\mu,A})\otimes \boldsymbol {\varepsilon}_{B}(\sigma_{\nu,B})]=\frac 14 \sum_{\mu,\nu=0}^{3}\Theta_{\mu\nu}(\sigma_{\mu,A}\otimes \sigma_{\nu,B}),
\end{equation}
where the part of $\boldsymbol {\varepsilon}_{\alpha}(\sigma_{\mu,\alpha}), \; \alpha=A,B$ is given by
\begin{align}
\label{eq:(12)}
\boldsymbol {\varepsilon}_{\alpha}(\mathbb{I}_{\alpha}) &= \mathbb{I}_{\alpha}+\sum_{\mu=1}^{3} \lambda_{\mu,\alpha}\cdot \sigma_{i,\alpha}\nonumber \\
\boldsymbol {\varepsilon}_{\alpha}(\sigma_{\nu,\alpha})   &= \sum_{\mu=1}^{3}\Gamma_{\mu \nu,\alpha}\cdot \sigma_{\mu,\alpha} \quad \nu=1,2,3.
\end{align}
The matrix $\boldsymbol{\Theta}$ for the evolved state is $\boldsymbol{\Theta}=\Big( \begin{matrix} 1 & \boldsymbol \upsilon_{B}^{\mathrm{T}} \\ \boldsymbol \upsilon_{A} & \boldsymbol S \end{matrix}\Big)$ and the vectors satisfy that $\boldsymbol \upsilon_{\alpha}=\boldsymbol \Gamma_{\alpha} \cdot \boldsymbol \upsilon_{\alpha}^{(0)}+ \boldsymbol \lambda_{\alpha}, \; \alpha=A,B$. According to Eqs. (11) and (12), the correlation matrix $\boldsymbol S $ for the evolved state is obtained by
\begin{equation}
\label{eq:(13)}
\boldsymbol S=\boldsymbol \lambda_{A}\cdot \boldsymbol \upsilon_{B}^{\mathrm{T}}+(\boldsymbol \upsilon_{A}-\boldsymbol \lambda_{A})\cdot \boldsymbol \lambda_{B}^{\mathrm{T}}+\boldsymbol \Gamma_{A}\cdot \boldsymbol T^{(0)}\cdot \boldsymbol \Gamma_{B}^{\mathrm{T}}.
\end{equation}
Here, $\boldsymbol {\varepsilon}_{B}=\boldsymbol  \varepsilon^{+}_{\mathrm{out}}$ is given by the quantum channel for Dirac particles from the outside region near the event horizon of the black hole.

To further explore the decoherence of Dirac particles, we study the dynamics of quantum coherence and quantum steering. The Werner state is chosen to be an initial one. In the Hilbert space spanned by $\{|00\rangle, |01\rangle, |10\rangle, |11\rangle \}$, $\rho_{AB}^{(0)}=\frac {1-\eta}4 \mathbb{I}+\eta |\psi^{-}\rangle \langle \psi^{-}|$ where $|\psi^{-}\rangle=\frac 1{\sqrt{2}}(|01\rangle-|10\rangle)$ and the parameter $\eta \in [0,1]$. In this case, the evolved state has a real $\mathrm{X}$ shaped form. According to quantum resource theory, the measure of quantum coherence based on $l_1$ norm is obtained by
\begin{equation}
\label{eq:(14)}
C(\rho_{AB})=\frac 12(|S_{11}-S_{22}|+|S_{11}+S_{22}|),
\end{equation}
where $S_{ij}$ are the elements of the correlation matrix for the evolved state. Recently, the inequality has been developed to judge whether a bipartite state is steerable when both of the observers are allowed to measure observations in their local sites \cite{Costa2016}. This inequality is composed of a finite sum of bilinear expectation values, which is given by $ F_n^{\mathrm{CJWR}}(\rho,\mu)=\frac{1}{\sqrt n}|\sum_{i=1}^n\langle A_i\otimes B_i\rangle|\leq 1$. Here, $A_i=\boldsymbol{u}_i\cdot\boldsymbol{\sigma},B_i=\boldsymbol{v}_i\cdot\boldsymbol{\sigma}$ where $\boldsymbol{u}_i\in \mathbb R^3$ are unit vectors and $\boldsymbol{v}_i\in \mathbb R^3$ are orthonormal vectors and $\boldsymbol {\zeta}=\{\boldsymbol {u}_1,\dots,\boldsymbol {u}_n,\boldsymbol{v}_1,\dots,\boldsymbol{v}_n\}$ is the set of measurement directions. The $\mathrm{CJWR}$ inequality for $n=2$ is used to quantify steerability $Q(\rho_{AB})=\max [0,\frac {F_2(\rho_{AB})-1}{F_2^{max}-1} ]$ where $F_2(\rho_{AB})=\max_{\boldsymbol{\zeta}}F(\rho_{AB},\boldsymbol{\zeta})$ and $F_2^{max}=\max_{\rho_{AB}} F_2(\rho_{AB})$. In this case, the steerability is obtained by
\begin{equation}
\label{eq:(15)}
Q_{s}(\rho_{AB})=\max [0,\frac {\sqrt{\mathrm{Tr}(\boldsymbol{S}\boldsymbol{S}^{\mathrm{T}})-\min_{i}(S^2_{ii})}-1}{\sqrt{2}-1} ].
\end{equation}
To clearly demonstrate the effects of external noises on the quantumness of Dirac particles, we consider a general open system model which is applicable to a qubit subjected to the fluctuating quantum fields with random phase noises. This kind of reservoir is modelled by the random telegraph noise channel. The noise is characterized by the autocorrelation function given as $\langle\chi(\tau)\chi(s)\rangle=\delta^2e^{-\gamma|\tau-s|}$ \cite{Daffer2004,Benedetti2013}. $\chi$ denotes the stochastic variable and $\delta$ represents the coupling strength between the system an reservoir. And $\gamma$ signifies the fluctuation rate of the noise. In the Kraus representation, the quantum channel for the random telegraph noise is given by $\boldsymbol {\varepsilon}(\rho)=\sum_i K_i \rho K_i^{\dag}$ where the Kraus operators are $K_1(\nu)=\sqrt{ \frac {1+\kappa(\nu)}2}\mathbb{I}$ and $K_2(\nu)=\sqrt{ \frac {1-\kappa(\nu)}2}\sigma_3$ and $\kappa(\nu)=e^{-\nu}[\cos (\mu \nu)+\frac 1{\mu}\sin (\mu \nu)]$ with $\mu=\sqrt{(\frac {2\delta}{\gamma})^2-1}$ and $\nu=\gamma \tau$. In the Bloch representation, we can rewrite the random telegraph noise channel by using the mapping matrix and vector,
\begin{equation}
\label{eq:(16)}
\boldsymbol {\Gamma }_{\mathrm{RTN}}=\mathrm{Diag} \left[ \kappa(\nu), \kappa(\nu), 1 \right], \quad \boldsymbol \lambda_{\mathrm{RTN}}= \boldsymbol{0},
\end{equation}
where the mapping matrix $\boldsymbol {\Gamma }_{\mathrm{RTN}}$ has the diagonal form. If the Dirac particle $A$ suffers from the dissipation of random telegraph noise, the correlation matrix $\boldsymbol{S}$ in the pauli basis is obtained by
\begin{equation}
\label{eq:(17}
\boldsymbol{S}=\mathrm{Diag} \left[-\eta\kappa(\nu)q_R \cos r, -\eta\kappa(\nu) q_R \cos r, -\eta |q_R|^2\cos^2 r  \right].
\end{equation}
In accordance with the results of Eqs. (14) and (15), The measures of quantum coherence and quantum steering are obtained by $C(\rho_{AB})=|\eta\kappa(\nu)q_R|\cos r$ and $Q_{s}(\rho_{AB})=\max \left[ \frac {F_2-1}{\sqrt{2}-1}, 0\right]$ where the quantity $F_2=\sqrt{2|\eta\kappa(\nu)q_R|}\cos r$. It is seen that both of them satisfy the relation $Q_s=\max \left[ \frac {\sqrt{2}C(\rho_{AB})-1}{\sqrt{2}-1}, 0 \right]$. The interplay between the external noise and quantum fluctuations near the event horizon of the black hole is numerically analyzed. It is seen that the maximum of quantum nonlocality can be attainable in the case of $\eta=1,q_R=1$. The dynamical behavior of quantum coherence with the reservoir evolution time $\tau$ and dilatonic parameter $D$ is illustrated in Figure 2. In the condition of weak couplings $(\frac {2\delta}{\gamma})^2<1$, the values of quantum coherence are monotonously decreased with the increase of the reservoir parameter in Fig. 2(a). The dilaton effect suppresses quantum coherence when the dilatonic parameter is large. In the strong coupling of $(\frac {2\delta}{\gamma})^2>1$, Figure 2(b) demonstrates that the revivals of quantum coherence oscillate with time. The result shows that the quantum fluctuation from the external reservoir leads to the oscillation of quantum coherence. The similar behavior of quantum steerability can be shown in Figure 3. In the evolution process, the values of quantum steerability are diminished to zero in the weak coupling case. While the death and revival of the steerability are presented in the strong coupling condition. These results demonstrate that the evolutions of quantum coherence and steerability are dependent on the interplay between the external noise and quantum fluctuations near the event horizon of the dilatonic black hole. The non-monotonous decoherence effects from the external environment can protect quantum nonlocality against the information loss from the relativistic effects of the dilatonic black hole.

\section{Discussion}

The effects of black hole's dilaton and external reservoir fluctuation on the quantumness for Dirac particles are investigated in the unified quantum channels. The process of information loss from quantum fluctuation near the event horizon of the black hole is modelled by the quantum channel beyond single mode approximation. The geometric feature and quantity of quantumness for the channel are obtained. We find out that the quantumness for Dirac particles escaping from the outside region can be preserved to a high extent under the circumstance of $q_R \rightarrow 1$ and small dilaton. In order to explore the decoherence of quantum nonlocality, we also put forward a general method by using the Pauli basis. The interplay between the external environmental noise and relativistic effects of the dilatonic black hole can be further studied. For an instance of external reservoir, the random telegraph noise channel is chosen. In the weak coupling condition, the monotonous degradation of quantum coherence and quantum steering with time can be observed when the dilaton parameter is increased. The oscillation revivals of quantum coherence and steerability are also presented when the strong coupling is considered. From the viewpoint of relativistic quantum information, this kind of non-monotonous decoherence contributes to the protection of quantum resources from the information loss near the event horizon of the black hole.

\begin{acknowledgments}
We would like to thank Professor Bill Unruh and Professor Philip C. E. Stamp for the discussions on the related work. The authors were supported by Postgraduate Research and Practice Innovation Program of Jiangsu Province. This work is funded by $\mathrm{SCOAP}^{3}$.
\end{acknowledgments}

\newpage

{\bf  Figure Captions}

Figure 1: The quantumness of Dirac particles in the outside region from the event horizon is illustrated with the variation of the parameter $q_R$ and dilaton value $D$. The sphere contraction in the Bloch representation is pictured when $q_R=0.1$ and $D=0.95$. The contour plot of the quantumness quantity based on quantum coherence is shown. Some parameters of the dilatonic black hole $M=1,\; \omega=1$ are chosen.

\vskip 0.5cm

Figure 2: The evolutions of quantum coherence are plotted as a function of the dilaton $D$ and reservoir time $\tau$ for the parameters of $M=1,\; \omega=1$ and $\eta=1, q_R=1$. The left plot denotes the weak coupling case where the reservoir parameters $\delta=0.05,\gamma=1$ are chosen. The right plot represents the strong coupling case where $\delta=0.05,\gamma=0.001$ are chosen.

\vskip 0.5cm

Figure 3: The decoherence of quantum steering is illustrated for the parameters of of $M=1,\; \omega=1$ and $\eta=1, q_R=1$. The black dash line depicts the evolution of the steerability when the weak coupling $\delta=0.05,\gamma=1$ is considered. The solid red line corresponds to the case of the strong coupling when $\delta=0.05,\gamma=0.001$ are chosen.

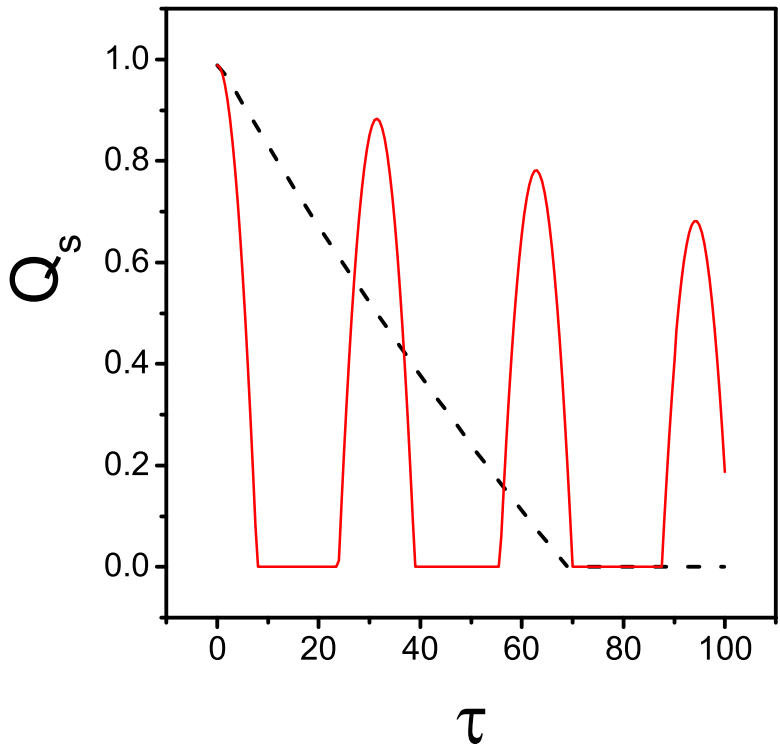
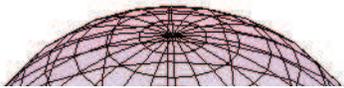
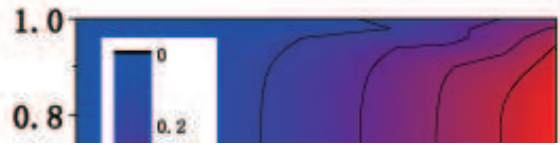
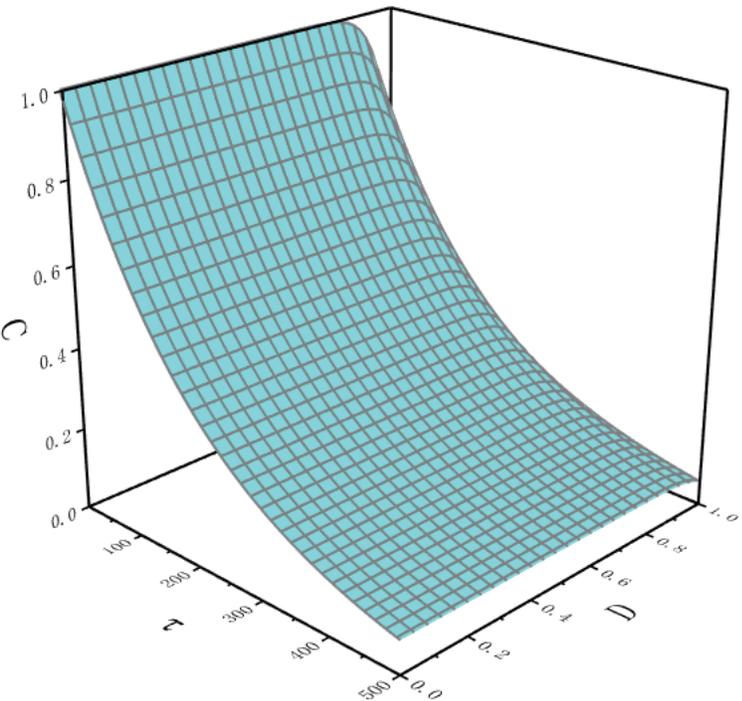
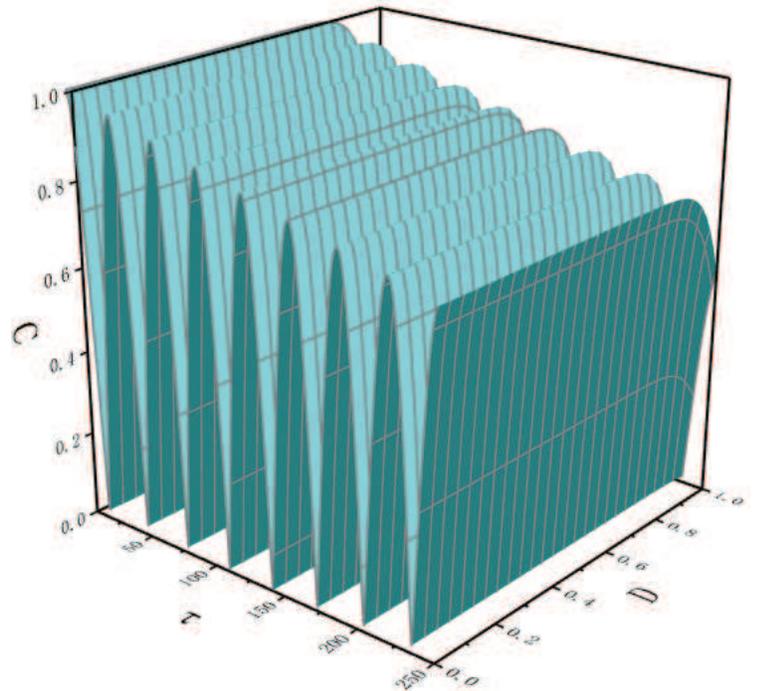
\end{document}